\documentclass[a4paper,amsmath,amssymb]{jpconf}
\usepackage{graphicx}
\usepackage{amsmath, amsthm, amssymb}
\begin{document}
\title{Causal analysis, Correlation-Response, and Dynamic cavity}

\author{Erik Aurell}

\address{Dept Computational Biology, ACCESS Linnaeus Centre
and Center for Quantum Materials, KTH-Royal 
Institute of Technology, SE-100 44 Stockholm, Sweden
\textit{and} Depts of Information and Computer Science and Applied Physics, 
Aalto University, P.O. Box 15400, FI-00076 Aalto, Finland}

\ead{eaurell@kth.se}

\author{Gino Del Ferraro}

\address{Dept Computational Biology, KTH-Royal 
Institute of Technology, SE-100 44 Stockholm, Sweden}

\ead{gino@kth.se}

\begin{abstract}
The purpose of this note is to point out analogies between causal analysis in statistics
and the correlation-response theory in statistical physics.
It is further shown that for some systems the dynamic cavity 
offers a way to compute the stationary state of a non-equilibrium process
effectively, which could then be taken an alternative starting point
of causal analysis.
\end{abstract}
\section{Causality in Philosophy, Physics and Statistics}
\label{sec:Introduction}
Causality formalizes the universal human experience of agents (causes) 
taking actions leading to results (effects)~\footnote{The literature on this topic
is too vast and variegated to be meaningfully referenced; two classic
studies from the perspective of belief systems in traditional societies
can be found in \protect\cite{LeviStrauss} and \protect\cite{Bourdieu}.}. In the Western Philosophical tradition
Aristotle postulated four kinds of causes namely
the material, the formal, the efficient and the final,
out of which Bacon retained the material and the efficient.
The material cause is in modern terms the material properties
of an object, ultimately determined by which atoms it is composed of
and how these are ordered in space,
while the efficient cause is how the object is influenced by other objects.
The Third Law of Newton~\cite{ThirdLaw} however states that in Nature 
there is no separation between cause and effect in the Aristotelian or Baconian sense;
Physics fundamentally knows only interactions, and these are always mutual,
a state of affairs unchanged since that time
and the replacement of the Classical Physics by Quantum Physics.
When the term causal is used to describe an interaction in modern high-energy Physics 
it means only that the influence cannot propagate faster than light so that
object $A$ at time $t_A$ only depends on what happened at object $B$ at times
$t_B$ early enough that a signal from $B$ can reach $A$ at time $t_A$, and \textit{vice versa}~\cite{Weinberg}.
  
The everyday and the philosophical notions of causality 
are in Physics instead intertwined with reversibility and irreversibility; Nature's laws
are time-reversal invariant on the fundamental level, but most ordinarily encountered processes
are overwhelmingly likely to only flow in one direction~\cite{Feynman,ChibbaroRondoniVulpiani}. We say that dropping a 
glass vase on the floor is the cause of it breaking because it is exceedingly unlikely that
the glass pieces would jump back together and fuse into a vase.
Similarly, we say an enzyme causes a chemical reaction in one direction when the concentrations
of the reactants are such that the opposite reaction is very unlikely.
Although the details of this complex process are not fully known, we can also 
say that smoking causes cancer because the DNA in living cells is mostly that of
one and the same genome for each individual -- an extremely small subset of
all possible DNA sequences of the same length -- and cancerogenes in tobacco smoke 
therefore almost always lead to mutations away from the healthy genotype and into one out of very many deficient genotypes. 
In Nature cause-effect relationships are thus but abbreviations for processes in physical systems
so strongly driven out of thermal equilibrium that they mostly only go one way.

Nevertheless, causal analysis is an important branch of statistics,
describing the effects of interventions 
and answering questions of the ``if-so-then-what?'' character~\cite{Pearl2009,Pearl-Causality}. 
Interventions are then taken to be outside Nature, typically ascribed to a human agent, and 
causality is thus distinct from statistical association studies. 
If person $X$ is holding a glass vase and person $Y$ trips him over,
then person $X$ is quite likely to fall and break the vase. However, we
cannot know this for sure without observing the event as person $X$ might for instance be much larger and stronger than $Y$.
Likewise, if we can deactivate enzyme $E$ then we can observe that a catalyzed reaction $S_1\to S_2$ ceases,
while if we can over-activate $E$ then the catalyzed reaction goes faster. This
is the paradigm for how molecular biologists identity interactions experimentally;
good research practice and common sense hold that observing such direct responses
is a more reliable means to acquire knowledge 
than observing the variations of
$E$ and the speed of the reaction $S_1\to S_2$ in natural undisturbed conditions.
For example, the catalyzed reaction may be one in a series of reactions
$S_1\to S_2\to S_3\to S_4\to\cdots$, catalyzed by enzymes $E,E_2,E_3,\ldots$ and
the living cell may regulate the production of all these enzymes by the availability
of the first substrate $S_1$~\cite{Alon}. In this case $E,E_2,E_3,\ldots$ would all vary
positively with the speed of the reaction $S_1\to S_2$ and only a direct
experiment can determine which of them actually does the job.

The first purpose of this paper is to show that there is a conceptual parallelism between 
causal analysis in statistics and long-time response functions in physics.
A main difference is that the results of an intervention in causal analysis
is (comparatively) easy to determine, while the response function integrates dynamical information
on all time scales. 
The second purpose is to show how the tools of message-passing/Belief Propagation,
developed to analyse complex static interdepencies, have been generalized to
describe dynamics with complex interactions, and then known as dynamic cavity.
Under suitable assumptions the dynamic cavity simplifies considerably the determination long-time responses,
which can hence be considered an alternative starting point of a causal analysis.
The paper is organized as follows. In Section~\ref{sec:do-calculus}
we describe in simple terms causal analysis, following
mainly~\cite{Do-Calculus}.
In Section~\ref{sec:correlation-response} we give a short summary of correlation-response theory
using Markov chains (synchronously updated spin systems) as our main example,
and show the parallelism between a causal analysis and correlation-response.  
In Section~\ref{sec:dynamic-cavity} we describe dynamic cavity as a generalization of message-passing/Belief Propagation to
dynamic phenomena, and show that it can be used to turn correlation-response into an alternative to causal analysis, for some systems.
In Section~\ref{sec:discussion} we sum up and discuss our results.

\section{Causal analysis}
\label{sec:do-calculus}
Statistical physicists are nowadays conversant with graphical models to describe probability distributions~\cite{YedidiaFreemanWeiss,MezardMontanari}.
To explain causal analysis we will start with two very simple Bayesian belief networks
\begin{equation}
\begin{array}{llcrr} 
(a)\qquad & \put(5,3){\circle{13}} A \leftarrow \put(5,3){\circle{13}}B \rightarrow \put(5,3){\circle{13}}C 
          & \hbox{and} &  (b)\qquad & \put(5,3){\circle{13}}A \rightarrow \put(5,3){\circle{13}}B \rightarrow \put(5,3){\circle{13}}C
\end{array}
\label{eq:2-BBNs}
\end{equation}
where by $\put(5,3){\circle{13}}A \rightarrow \put(5,3){\circle{13}}B$ we mean that both the random variable $B$ is dependent on random variable $A$
in the ordinary sense of probability, and also that $A$ (somehow) causes $B$.
The dependency is encoded in conditional probabilities $P_{A|B}(a|b)$ where $a$ and $b$ are
values of $A$ and $B$. The joint probabilities of the three variables are in the two cases
\begin{eqnarray}
\hbox{Case \textit{(a)}}\quad P_B(b) P_{A|B}(a|b) P_{C|B}(c|b) &=& \frac{P_{A,B}(a,b) P_{C,B}(c,b)}{P_B(b)} \nonumber \\
\hbox{Case \textit{(b)}}\quad P_A(a) P_{B|A}(b|a) P_{C|B}(c|b) &=& \frac{P_{B,A}(ba) P_{C,B}(c,b)}{P_B(b)}  \nonumber
\end{eqnarray}
where $P_B(b)$ is the marginal probability of variable $B$ to take value $b$ while
$P_{A,B}(a,b)$ is the  marginal probability of the pair of variables $A$ and $B$ to take values $a$ and $b$, and so on. 
We assume for simplicity that $P(b)$ is different from zero for all values $b$ of $B$. 
Both Bayesian belief networks encode the same joint probability; we 
cannot distinguish by covariation whether $A$ causes $B$ or $B$ causes $A$. 
In the language of factor graphs
this joint probability can alternatively be described by a factor graph~\cite{Kschischang2001}
\begin{equation}
\begin{array}{c}  

\put(4,4){\circle{13}}A   
   \put(4,4){\line(1,0){15}}
      \put(20,0){\fbox{f}} 
         \put(30,4){\line(1,0){15}}
            \put(52,4){\circle{13}} \put(48,0){B}
               \put(60,4){\line(1,0){15}}
                  \put(75,3){\fbox{g}}
                     \put(88,4){\line(1,0){15}}
                        \put(110,4){\circle{13}}\put(106,0){C}   
 
            \put(52,-2){\line(0,-1){15}} 
              \put(47,-28){\fbox{h}} 


\end{array} 
\label{eq:factor-graph} 
\end{equation}
with factors 
\begin{equation}
f(a,b)=P(A=a,B=b)\qquad g(c,b)=P(C=c,B=b) \qquad h(b)= (P(B=b))^{-1}
\end{equation} 
and
\begin{equation}
P_{A,B,C}(a,b,c) = \frac{1}{Z} f(a,b) g(c,b) h(b)
\label{eq:joint-probability} 
\end{equation}
The partition function $Z$ is determined by the
interactions encoded by the factors of the factor graph,
and is in general difficult to compute for large models,
but in the simple example considered here it is obviously equal to one. 

In both cases described above the joint probability of $A$ and $C$ conditioned on $B$ is then
\begin{equation}
P_{A,C|B}(a,c|\hbox{see}(B=b)) = \frac{P_{A,B}(a,b) P_{C,B}(c,b)}{(P_B(b))^2}=P_{A|B}(a|b) P_{C|B}(c|b)
\label{eq:see}
\end{equation}
where we have introduced Pearl's ``see'' notation~\cite{Do-Calculus}.
To avoid confusion, let us note again that $P_{A|B}(a|b)$ in above has its ordinary probabilistic meaning of
$\frac{\sum_{c}P_{A,B,C}(a,b,c)}{\sum_{a,c}P_{A,B,C}(a,b,c)}$ and is the same in both models.

If we intervene on $B$ and set its value to $b$ the two Belief networks
lead to new joint probabilities on the two remaining variables
($A$ and $C$). In both cases $A$ and $C$ become independent with
probabilities depending parametrically on the set value $b$, 
which we can call $P_A^{(B)}(a;b)$ (``the probability distribution of random
variable $A$ in the modified model where variable $B$ has been set to constant $b$'')
and
$P_C^{(B)}(c;b)$ (``the probability distribution of random
variable $C$ in the modified model where variable $B$ has been set to constant $b$'').
Introducing Pearl's ``do'' notation~\cite{Do-Calculus} we then have 
\begin{equation}
P_{A,C}(a,c|\hbox{do}(B=b)) = P_A^{(B)}(a;b) P_C^{(B)}(c;b) 
\label{eq:do}
\end{equation}
and the dependencies can be illustrated as a (trivial) Bayesian network 
\begin{equation}
\begin{array}{c} 
   \put(0,4){\circle{13}} \put(-4,0){A}

   \put(100,4){\circle{13}} \put(96,0){C}
\end{array}
\label{eq:BBN-reduced}
\end{equation}
or as an (equally trivial) factor graph
\begin{equation}
\begin{array}{c} 

   \put(0,4){\circle{13}} \put(-4,0){A}
     \put(0,-2){\line(0,-1){15}} 
       \put(-5,-25){\fbox{e}} 

   \put(100,4){\circle{13}} \put(96,0){C}
     \put(100,-2){\line(0,-1){15}} 
       \put(95,-28){\fbox{d}}

\end{array}
\label{eq:factor-graph-reduced}
\end{equation}
The two probability distributions are however not the same in the two cases.

In case \textit{(a)} we have as numerical values $P_A^{(B)}(a;b)=P_{A|B}(a|b)$
and $P_C^{(B)}(c;b)=P_{C|B}(c|b)$, because $A$ and $C$ are then both assumed to be caused by $B$;
the factors in the factor graph (\ref{eq:factor-graph-reduced}) are $e=P_{A|B}(a|b)$ and $d=P_{C|B}(c|b)$.
For this case (\ref{eq:see})  and (\ref{eq:do}) hence describe the same distribution.

In case \textit{(b)} we also have
$P_C^{(B)}(c;b)=P_{C|B}(c|b)$, but for the other probability instead 
$P_A^{(B)}(a;b)=P_{A}(a)$ corresponding to a factor $e=P_{A}(a)$ in (\ref{eq:factor-graph-reduced}).
This difference is ultimately what it means to interpret the arrows in
(\ref{eq:2-BBNs}) as causes: if $A$ is a cause and $B$ is an effect then $A$ should be  
unaffected by $B$, and in particular unaffected by any outside intervention on $B$.
Therefore, whether or not there is any intervention on $B$, in case \textit{(b)} the marginal probability 
of $A$ is and remains $P_{A}(a)$ and the ``do'' (\ref{eq:do}) is different from the ``see'' (\ref{eq:see}).
Expanding on the same point, in case \textit{(b)} the ``do-probability'' $P_{A,C}(a,c|\hbox{do}(B=b))$
can be expressed in terms of probabilities observable before the intervention,
namely as $P_{A}(a) P_{C|B}(c|b)$, but is not the same
as the ``see-probability''  $P_{A,C}(a,c|\hbox{see}(B=b))$ which is, for both cases, $P_{A|B}(a|b)P_{C|B}(c|b)$.
The Kullback-Leibler distance between the two is (for this simple example) $KL(\hbox{see}(B=b)|\hbox{do}(B=b))=
\sum_aP_{A|B}(a|b)\log \frac{P_{A|B}(a|b)}{P_A(a)}$, which generally is not zero.

\begin{figure}[ht]
\begin{center}
\includegraphics[width=7.9cm]{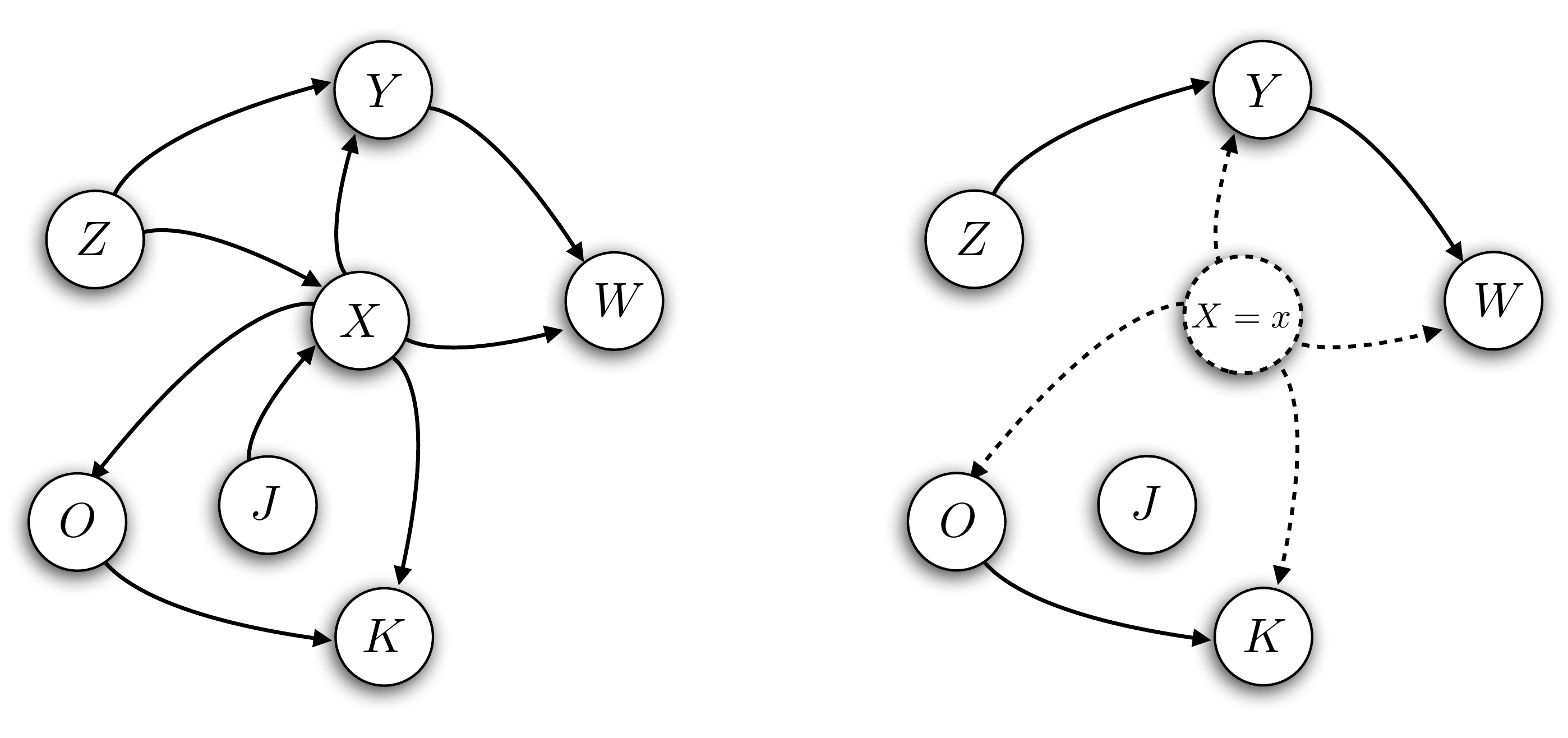}
\end{center}
\caption{\footnotesize 
Illustration of the do operation. Left panel: a Bayesian belief network with a central node containing variable $X$.
Right panel: reduced Bayesian belief network after intervening on variable $X$ setting it to value $x$. Node containing $X$
and outgoing links are indicated by dashed lines symbolizing that (\protect\ref{eq:joint-probability-post}) depends parametrically on $x$.
Ingoing links to node containing $X$ are eliminated together with random variable $X$ which does not appear in (\protect\ref{eq:joint-probability-post}).
}
\label{fig:do-see}
\end{figure}

We will now raise the abstraction level and following~\cite{Do-Calculus}
define a general causal model $M$, also known as \textit{Structural Equation Model},
as a set of exogenous variables $U$, a set of endogenous variables 
$V_1,\ldots,V_N$, located in nodes $1,\ldots,N$ in a graph $G$, 
for each node $i$ a set of parent nodes $PA_i\subset \{ \{1,\ldots,N\}\setminus i\}$
and conditional probabilities $F_i(V_i|V_{PA_i}, U)$.
The structure of $G$ is determined by there being a link $i\to j$ iff $i\in PA_j$.
Additionally one may include in the model specifications a distribution $P(U)$ over the exogenous variables~\cite{Do-Calculus}.
Each such model defines a joint probability distribution of the endogenous variables as
\begin{equation}
P_M(V_1,\ldots,V_N| U)=\frac{1}{Z_M(U)}\prod_i F_i(V_i|V_{PA_i}, U)
\label{eq:joint-probability-general} 
\end{equation}
If $G$ is a Directed Acyclic Graph (DAG), so that dependencies cannot propagate in a loop, clearly $Z_M(U)=1$.
The \textit{do operator} is introduced by Pearl as:
\begin{quote}
Interventions and counterfactuals are defined through a mathematical operator called $do(x)$, which simulates physical 
interventions by deleting certain functions from the model, replacing them with a constant $X=x$, while keeping the rest of the model unchanged. 
The resulting model is denoted $M_x$. The postintervention distribution resulting from the action $do(X=x)$ is given by the equation
\begin{equation}
P_M(y|do(x)) =P_{M_x}(y) \nonumber
\end{equation}
\begin{flushright}  
Judea Pearl, ``The Do-Calculus Revisited'' (2012)~\cite{Do-Calculus}
\end{flushright}    
\end{quote}
It is useful to compare and contrast the do operation with the cavity method to be discussed in more detail below in Section~\ref{sec:dynamic-cavity}.
Both modify a probabilistic model by eliminating one or more variables, figuratively opening a hole (or cavity) in the factor graph.
A first difference is that in the cavity method the variable and all its interactions are eliminated as if they were never there,
while in a do operation the variable is set to a constant and the value of that constant matters. 
Instead of (\ref{eq:joint-probability-general}) we thus have, taking $X=V_k$ for some $k$,
\begin{equation}
P_{M_{v_k}}=\frac{1}{Z_{M_{v_k}}(U)}\prod_{i\neq k \atop k \not\in V_{PA_i}} F_i(V_i|V_{PA_i}, U)
\prod_{i\neq k \atop k \in V_{PA_i}} F_i^{V_k=v_k}(V_i|\{V_{PA_i}\}\setminus V_k, U)
\label{eq:joint-probability-post} 
\end{equation}
where $Z_{M_{v_k}}$ is a new normalization constant and $F_i^{V_k=v_k}$ is a new function obtained from $F_i$ by setting the variable
$V_k$ to the constant $v_k$. Hence we can express the ``see'' and the ``do'' as
\begin{equation}
P(V_i|\hbox{see}(V_k=v_k)) = \frac{P_M(V_i,V_k)}{P_M(V_k)} \qquad\qquad P(V_i|\hbox{do}(V_k=v_k)) = P_{M_{v_k}}(V_i)
\label{eq:see-and-do-in-general} 
\end{equation}
which again shows how and why the two concepts differ. A graphical illustration of the do operation given in Fig.~\ref{fig:do-see}.

A second and more important difference is that the do operation is formulated for Bayesian belief networks 
of which (as we have seen) there can be many corresponding to the same joint probability distribution.
Under the operations $\hbox{do}(X)$ for different $X$, each Bayesian belief network (\textit{i.e.}
each direction of the arrows in Eq.~(\ref{eq:2-BBNs})) hence specifies 
a different set of changes of the joint probability distribution encoded in a factor graph.

An important question in causal analysis is whether probabilities after an intervention, 
\textit{i.e.} $P_M(y|do(x))$, $y$ standing for any subset of the endogenous variables,
can be determined from observations before the intervention, \textit{i.e.} from the set $P_M(z)$ (any and all $z$).
When this is so one says that a causal effect query is \textit{identifiable} 
because it can be decided (the probability $P_M(y|do(x))$ estimated) from
data obtained before an intervention.
In both the simple examples above this was the case, only the $P_M(y|do(x))$'s were not the same.
More generally, a causal effect query is always identifiable from passively observed $P_M(z)$, provided that all variables in $M$
are observed and $G$ is known~\cite{Pearl-Causality}.
In less technical terms this last statement means nothing else than given sufficient data one can in principle 
estimate conditional probabilities, and given a direction of the arrows in a Bayesian belief network one can translate
this information into what the conditional probabilities will be in a modified model; the situation is more complicated when 
some variables are unobserved (un-measured). The \textit{Do-Calculus} of Pearl consists of three rules (a calculus, shown to be complete) 
for deciding identifiability when $G$ is known and is a DAG, and some of the variables are unobserved. 
The Do-Calculus can hence be used to determine (in perhaps quite complex settings)
whether a separate experiment is necessary, or if an hypothetical question can be
answered with the data already at hand.
Other uses of Do-Calculus are \textit{mediation}, how to separate direct and indirect
causal relationships, \textit{transportability}, how to determine how much information
can be carried over to a new condition, and \textit{meta-analysis},  
how to pool data obtained under different conditions, for more details
see~\cite{Do-Calculus}.

\section{Correlation-Response}
\label{sec:correlation-response}
At the basis of scientific mathematical philosophy is the idea 
that the regularities of the world are best expressed by how it
changes in time, famously stated by Newton to Leibniz as   
\textit{6accdae13eff7i319n4o4qrr4s8t12ux}~\cite{Arnold-Newton}.
Let us therefore substitute the Bayesian belief network in Section~\ref{sec:do-calculus}
by a minimal model encoding the same dependencies as a probabilistic evolution law: 
\begin{equation}
P(V_1(t),\ldots,V_N(t))=\prod_{i=1}^N F_i(V_i(t)|V_{PA_i}(t-1)) P(V_1(t-1),\ldots,V_N(t-1))
\label{eq:joint-probability-dynamic} 
\end{equation}
The notation is here the same as in (\ref{eq:joint-probability-general}) except that the variables are now
indexed by time ($t$) and a possible dependence on exogenous variables has been suppressed.
Up to the technical simplification of synchronous dynamics, (\ref{eq:joint-probability-dynamic}) is a prototype for a physically
realistic mutual dependency. It could be realized in a biological
regulatory system, say in a signal transduction network, where the cause-effect relationship between
$PA_i$ and $V_i$ would have the underlying mechanistic interpretation 
of $PA_i$ being the kinases, phosphatases and other enzymes catalyzing the phosphorylation, de-phosphorylation and other modifications to unit $i$.
The endogenous variables $U$ are then concentrations of molecules at constant concentrations, which could be sugars and other carbon sources
for bacteria, or hormones and other signalling molecules in multi-cellular organisms.  
For long times the probability distribution in (\ref{eq:joint-probability-dynamic})
would then reach stationary state which we will denote
\begin{equation}
P^*(V_1,\ldots,V_N| U)=\lim_{t\to\infty}P(V_1(t),\ldots,V_N(t)| U)
\label{eq:limit}
\end{equation}
Since $P^*$ in (\ref{eq:joint-probability-dynamic}) is at least as realistic
as $P$ in (\ref{eq:joint-probability-general}) as a representation of how
the endogenous variables depend on the exogenous variables we
could also use it to define an analogy of the do operation.
We thus have 
\begin{equation}
P_M^*(y|\hbox{do}(x=X))=P_{M_x}^*(y) = \lim_{\tau\to\infty}R(y,x,t+\tau,t)
\label{eq:do-in-the-limit}
\end{equation}
where the last equality says that this is simply the long-time limit of a response function.

Generally a response function related to a generic quantity $V_i(t)$ is 
defined as $ R_{ij}(t,t') = \langle \partial V_i(t)/ \partial H_j (t') \rangle$ where $H_j(t')$ is a general parameter which can be varied within the system.  
An example, in ferromagnetic systems, is represented by the susceptibility function 
$\chi_{ij}(t,t') = \langle \partial M_i(t)/ \partial H_j(t') \rangle$ which gives a measure of the change in the local magnetization $M_i(t)$ on site $i$ 
at time $t$ due to a change of an external field $H_j(t')$ acting on a different site $j$ at an earlier time $t'$. 
In this respect, the \emph{do} operation, otherwise stated as an intervention on the system, is similar to the variation of an (or several) 
external parameter(s) in a physical system which generates a response function. 
A main goal of causal analysis can be then reformulated as predicting suitable defined response functions under the change of 
some external tuneable parameters (intervention). 
A great deal is known about such response functions for systems at or near thermodynamic equilibrium
where they are related to correlation functions through the
Fluctuation-Dissipation-Theorem, which generically takes the form~\cite{Kubo,Villamaina} 
\begin{equation}
\frac{1}{T}[C(\tau=0)- C(\tau)] = \int_0^\tau R(\tau') d \tau'
\end{equation}
where $\tau= t-t'$, time translational invariance is assumed, and $T$ is temperature.  
We note that the left hand side of above is measured in the unperturbed system and the right hand side in the perturbed system. 
Causal effect queries are therefore always identifiable in systems at or near thermodynamic equilibrium
from observing no more than the correlation between the variable which is set and the variable one wants to predict.
The relation between correlation and response has been also used to improve for inference on a network~\cite{Raymond}.

\section{Dynamic cavity}
\label{sec:dynamic-cavity}
In this section we describe how the techniques now generally called message-passing or Belief Propagation 
can be generalized to analyse evolution laws like (\ref{eq:joint-probability-dynamic}).
Message-passing techniques have been invented independently in different fields~\cite{YedidiaFreemanWeiss,Kschischang2001}.
In Physics they are also known as the cavity method~\cite{MezardMontanari}, and usually traced back to~\cite{Bethe1935}.
Their purpose is to compute marginal probabilities over some (usually small) subset of
variables in a probabilistic model described by a factor graph which is done by storing partial computations
in nodes representing the variables and then forwarding such partial results to neighbours in the graph
for further processing.  
Message-passing converges and is exact if the underlying graph is a tree
but also often converges and is a very good approximation if the underlying graph has only long loops, 
a fact that has many theoretical and practical applications in coding theory and elsewhere~\cite{MezardMontanari}.
The fixed points of the algorithms correspond to stationary points under variation
of the Bethe approximation to the free energy in the corresponding statistical mechanics problem
\cite{YedidiaFreemanWeiss,MezardMontanari}.

Our point of departure is now the observation that the dynamics (\ref{eq:joint-probability-dynamic}) naturally 
leads to a probability distribution on variable histories
\begin{equation}
P(X_1,\ldots,X_N)=P(V_1(0),\ldots,V_N(0))\prod_{t=1}^T \prod_{i=1}^N F_i(V_i(t)|V_{PA_i}(t-1))
\label{eq:joint-probability-history} 
\end{equation} 
where $X_i=\{V_i(0),\ldots,V_i(T)\}$ for $i=1,\ldots,N$.
Before continuing, let us note that if the variables are Boolean and take values $\{-1,1\}$ (``spins'')
then (\ref{eq:joint-probability-dynamic}) specifies a dynamics of a spin system under synchronous updates,
and if further all the transition probabilities are of the type
$F_i(V_i(t)|V_{PA_i}(t-1))\propto \exp\left(V_i(t)\left(h_i+\sum_j J_{ij} V_j(t-1)\right)\right)$
is known as the Kinetic Ising model~\cite{Glauber}. The parent set $V_{PA_i}$ is then comprised of the variables $V_j$
for which $J_{ij}$ is non-zero.
When $J_{ij}=J_{ji}$ for all pairs $(i,j)$ the system has a stationary state 
$P(V_1,\ldots,V_N)\propto \exp\left(\sum_i h_i V_i+\sum_{ij} J_{ij} V_i V_j\right)$
and (\ref{eq:joint-probability-dynamic}) then simulates a system in thermal equilibrium, albeit under the somewhat
unphysical synchronous update rule.
In the more general case when  $J_{ij}\neq J_{ji}$, and in particular for fully asymmetric models where 
$J_{ij}$ can only be non-zero when $J_{ji}$ equals to zero, (\ref{eq:joint-probability-dynamic}) on the other hand simulates a non-equilibrium system.

The first result on reducing the complexity of (\ref{eq:joint-probability-history}) dates back almost thirty 
years~\cite{Derrida1987} and pertains to fully asymmetric models.
For these an influence $X_j\to X_i$ must traverse a loop in $G$ to get back to $X_j$, and when there are no loops,
or when these can otherwise be disregarded, the marginal probability of $X_j$ is independent of $X_i$.
This leads to simple equation for the marginalization over a single variable and and single time, namely
\begin{equation}
P_i(V_i,t)= \sum_{V_j\in PA_i} F_i(V_i|V_{PA_i})\prod_{j} P_j(V_j,t-1)\qquad\hbox{(Fully asymmetric})
\label{eq:Derrida} 
\end{equation}   

We now generalize a bit and assume that the dependency graph $G$
has the property associated with the effectiveness of standard message-passing
\textit{i.e.} that it is a tree, or at least locally tree-like. That is,
we assume that one cannot form circular dependency chains
$i\to j\to k\to\cdots\to i$ where $V_i\in PA_j, V_j\in PA_k,\ldots\in PA_i$
unless either somewhere the chain backtracks as $\cdots j\to k\to j \to\cdots$
or the chain is long, on the order of the graph diameter of $G$.
The term ``dynamic cavity'' was introduced in~\cite{Kanoria2011} for such situations
where it was used to obtain rigorous bounds on the consensus threshold for the majority dynamics.
Methods have later been developed to treat, in principle exactly, such problems when
the dynamical law is modified to only allow transitions in one direction~\cite{lokhov2014dynamic,Altarelli2014}.
An important step was taken in~\cite{Neri2009} where marginals 
in a stationary state were computed approximately based on an ansatz,
recently extended to to a perturbative scheme and to
also cover transient phenomena~\cite{DelFerraro2015,BarthelDeBaccoFranz2015}.
The main problem is then that even when the dependency graph $G$ of the dynamics
in (\ref{eq:joint-probability-dynamic}) is locally tree-like this is not the case
for dependencies in (\ref{eq:joint-probability-history}) due to ``loops-in-time''.
These dependencies have been resolved by a graph expansion technique~\cite{lokhov2014dynamic,Altarelli2014,DelFerraro2015}
which we now explain.

\begin{figure}[ht]
\begin{center}
\includegraphics[width=7.9cm]{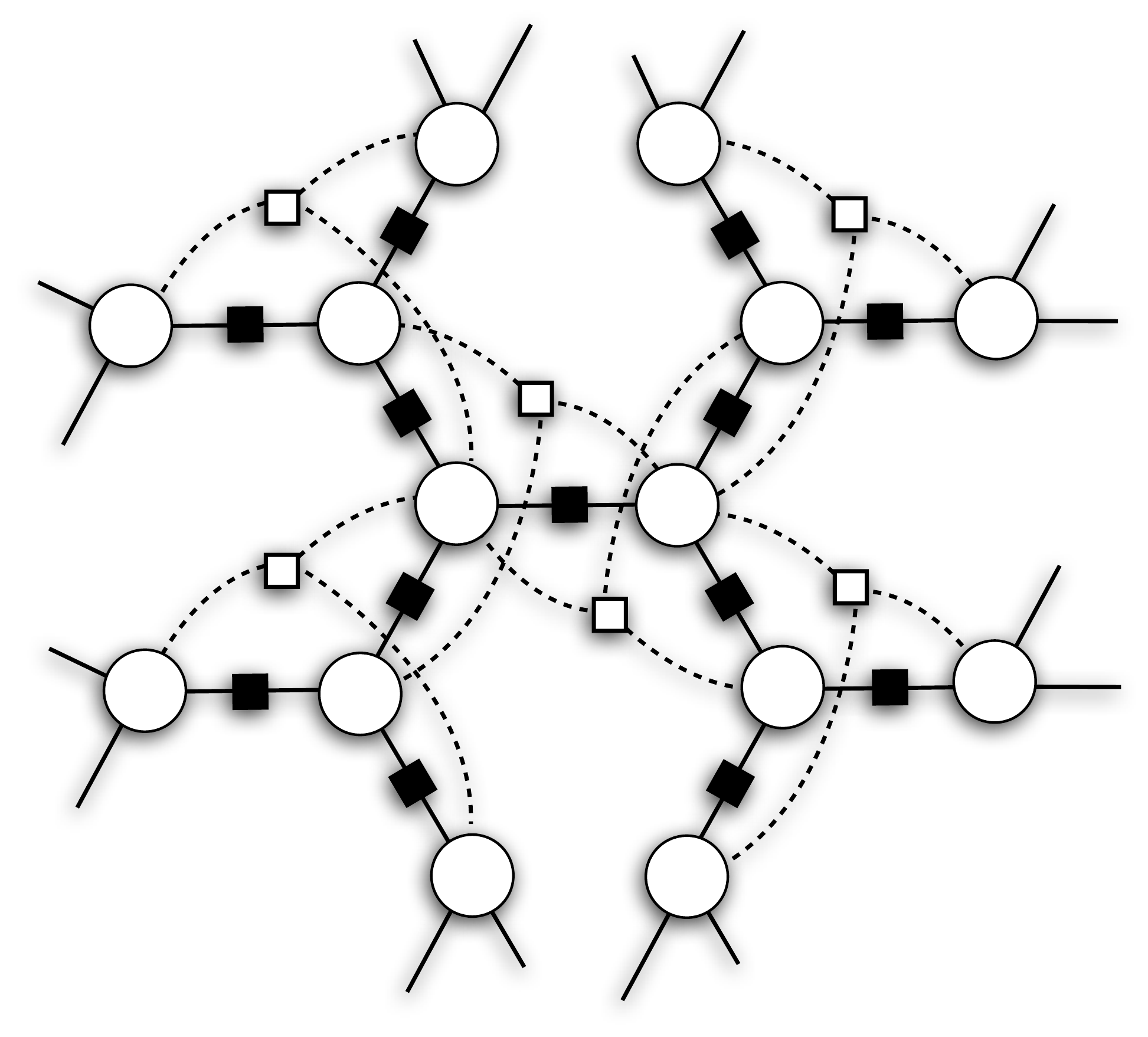}
\end{center}
\caption{\footnotesize 
A tree-like dependency graph with the normalization constants in the transition functions split off as separate
factor nodes (boxes). It has been assumed that the dependencies are not fully asymmetric so
that when node $i$ depends on node $j$, node $j$ in general also depends on node $i$.
Dependencies between nodes (in general mutual) are indicated by (undirected) lines.
In the kinetic model loops emerge from variables at different times in (\protect\ref{eq:joint-probability-history}).
}
\label{fig:fg1}
\end{figure}

First, as for the Kinetic Ising model it is often convenient to define transition
functions only up to a normalization  
$F_i(V_i|V_{PA_i})\propto \exp\left(r_i(V_i,V_{PA_i})\right)$.
The normalization constant is then $N_i(V_{PA_i})=\sum_{V_i}\exp\left(r_i(V_i,V_{PA_i})\right)$,
a function that does not depend explicitly on $V_i$.
Assuming further for simplicity that interaction functions $r_i$ are only pair-wise the dependency graph
can be illustrated as in Fig.~\ref{fig:fg1}.
The model defined on variable histories, (\ref{eq:joint-probability-history}), now has short-loop dependencies
even if the graph $G$ itself does not. This can be seen by tracing the dependency of one of these variable, say $X_i=\{V_i(0),\ldots,V_i(T)\}$.
Pick a time $t$ and notice that $V_i(t)$ depends on $V_j(t-1)$ for all $j\in PA_i$. Then pick two of these variables $X_j$ and  $X_k$
such that $i\in PA_j$ and  $i\in PA_j$, then $V_j(t-1)$ and $V_k(t-1)$ both depend on $V_i(t-2)$. At the same time 
$V_j(t-1)$ and $V_k(t-1)$ are however also dependent through the normalization $N_i(V_{PA_i})$, and  $X_i$, $X_j$ and $X_k$
are therefore connected in a dependency loop of length three.

\begin{figure}[ht]
\begin{center}
\includegraphics[width=7.9cm]{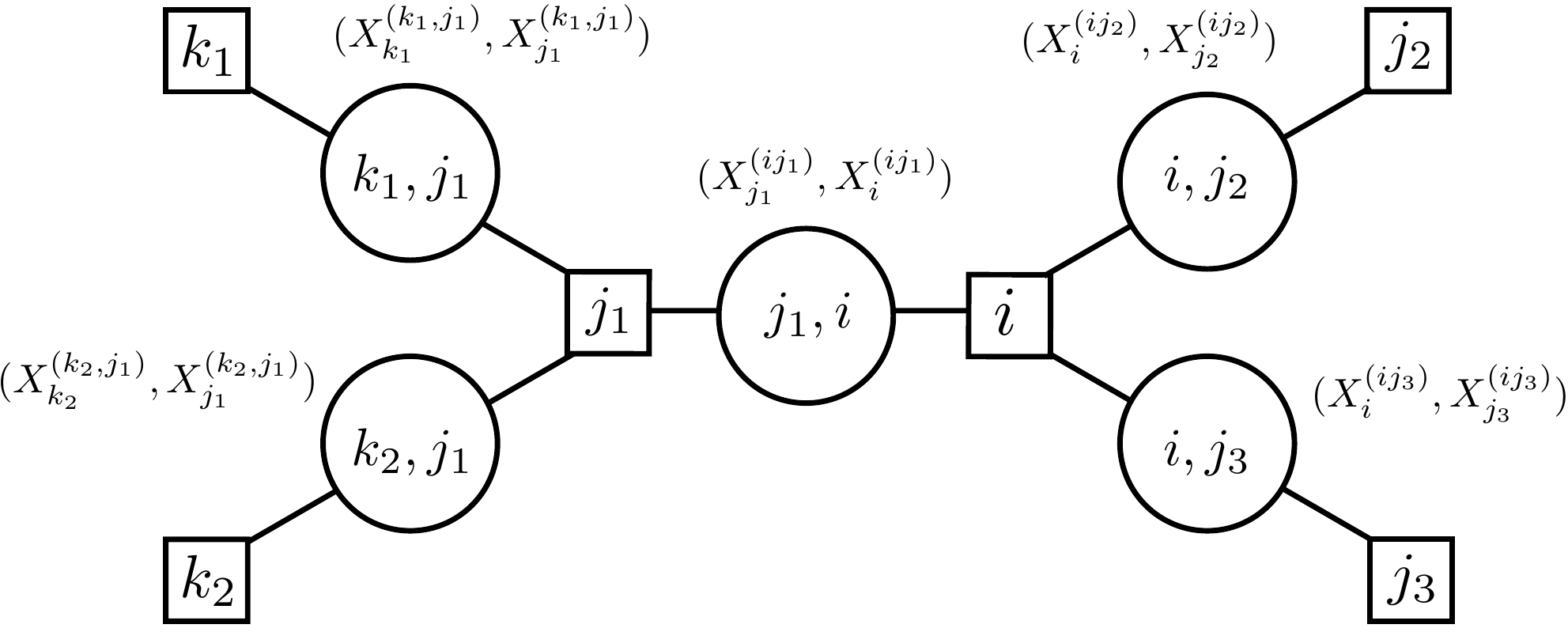}
\end{center}
\caption{\footnotesize 
Auxiliary graph obtained from a part of Fig.~\protect\ref{fig:fg1} where loops have been removed by a graph expansion procedure.
The new variable nodes contain histories of two variables that were neighbours in the original graph 
while new factor nodes (one per each old variable node) contain both the transition functions and the consistency conditions.  
}
\label{fig:fg2}
\end{figure}

While there are many approaches to get rid of loops in factor graphs we will use one
which is well adapted to the dynamics.
For every pair $i$ and $j$ such that $V_i\in PA_j$ (whether or not also  $V_j\in PA_i$)
we introduce a new compound variable $(X^{(ij)}_i,X^{(ij)}_j)$ interpreted as
``variable $X^{(ij)}_i$ of type $X_i$ belonging to link $(i,j)$ and 
 $X^{(ij)}_j$ of type $X_j$ also belonging to link $(i,j)$''~\footnote{To be precise the two parts of the
compound variable are distinguished by their index ($i$ or $j$) and not by their order in the pair.
When a message is to be transmitted from $i$ to $j$ they are naturally read in the
order $(X^{(ij)}_i,X^{(ij)}_j)$ while if the message is transmitted in the opposite
direction the natural order is $(X^{(ij)}_j,X^{(ij)}_i)$.}.
Introducing now the consistency requirement that the variables $X^{(ij)}_i$ take the same value for
all the links $(i,j)$ where this type of variable is found we can rewrite
(\ref{eq:joint-probability-history}) as
\begin{eqnarray}
P(\{X^{(ij)}_i,X^{(ij)}_j\})&=&P_{\hbox{init}} \prod_{t=1}^T \prod_{i} F_i(\overline{V_i(t)}|\{V^{(ij)}_j(t-1)\}_{j\in{PA_i}})  \nonumber \\
                          &=& \prod_{t=1}^T \prod_{i} \mathbf{1}_{V^{(ij_1)}_i(t)=V^{(ij_2)}_i(t)=\ldots} 
\label{eq:joint-probability-history-expanded} 
\end{eqnarray} 
where $P_{\hbox{init}}$ is the probability distribution on the initial conditions translated 
to the new variables, $V^{(ij)}_j(t)$ are the restrictions of the variables
$X^{(ij)}_j$ to a single time $t$
and $\overline{V_i(t)}$ is any suitable average of the $V^{(ij)}_i(t)$ for different $j$~\cite{DelFerraro2015}.
The graph expansion is illustrated in Fig.~\ref{fig:fg2}. 
Introducing messages in the standard way and summing out the consistency conditions we thus arrive at~\cite{DelFerraro2015}
\begin{equation} 
m_{i \to (ij)} (X_i^{(ij)}, X_j^{(ij)}) \propto \sum_{\{ X_k^{(ik)} \}}  \Phi_i(X_i^{(ij)}, X_j^{(ij)}, \{ X_k^{(ik)}\} ) 
\prod_{k \in \partial i \backslash j} m_{k \to (ik)} (X_k^{(ik)}, X_i^{(ij)})
\label{eq:dynamicBP} 
\end{equation}
where $\Phi_i(X_i^{}, X_j^{},\{X_k^{}\})=\prod_{t=1}^{T} F_i(V_i(t)|V_j(t-1),\{V_k(t-1)\}_{k\in{PA_i}\setminus j})$.
Equation (\ref{eq:dynamicBP}) are the dynamic cavity update equations
which are the same as ordinary cavity update equations applied to the model 
(\ref{eq:joint-probability-history}) on variable histories.
A trace of the dynamic origin remains in that the probability
$m_{i \to (ij)} (X_i^{(ij)}, X_j^{(ij)})$ can be taken to depend on the full history
$X_i^{(ij)}=\{V_i^{(ij)}(0),\ldots,V_i^{(ij)}(T)\}$ of the first argument, but only a one unit shorter
history $X_j^{(ij)\prime}=\{V_j^{(ij)}(0),\ldots,V_j^{(ij)}(T-1)\}$ of the second argument.
For a discussion as well as a description of the analogous dynamic cavity output equations, see~\cite{DelFerraro2015}.
To make (\ref{eq:dynamicBP}) practical further assumptions are needed, to close the 
iterations in a low-dimensional subspace of the functions $m_{i \to (ij)} (X_i^{(ij)}, X_j^{(ij)})$.
In~\cite{DelFerraro2015} good results were reported based on
closure in the class of 1-step Markov processes, leading to schemes
not much more complicated than (\ref{eq:Derrida}) while in~\cite{BarthelDeBaccoFranz2015} even better results were reported 
from more involved procedure. The field is in active development and likely even better approximations will appear in the near future.

We will now take the point of view that the probabilities $P^*$
and $P_{M_x}^*$ in (\ref{eq:limit}) 
and (\ref{eq:do-in-the-limit}) are efficiently computatable
and ask what are the implications for causal analysis.
First, the assumption of synchronous updates is unrealistic in most natural systems
but certainly no more so than the assumption of instantaneous dependence made in 
(\ref{eq:joint-probability-general}). 
In most problems where an underlying mechanistic explanation is conceivable
``causes'' are ultimately to be interpreted as variables influencing transition rates,
and the simplest example of such dynamics is (\ref{eq:joint-probability-dynamic}).
In stationary state an underlying explanation, which one could call ``mechanistic causes'',
leads to a joint probability distribution $P^*$ with generally many more dependencies.
That is, there will be one (directed) dependency graph 
$G$ describing the probabilistic evolution law  (\ref{eq:joint-probability-dynamic})
and another (undirected) factor graph $F$ describing the probability $P^*$ in (\ref{eq:limit}),
and $F$ will almost always be (much) larger and (much) richer than $G$. For a worked-out example
of such an effect, in the relaxation towards equilibrium of the Kinetic Ising model on a 1D 
lattice~\cite{Glauber}, see~\cite{DelFerraro2014}.

The stark conclusion of these in essence elementary
considerations is hence simply that causal analysis should be treated with a great deal of caution. 
The causes identified by causal analysis may not be causes at all in the everyday understanding of the word.
Instead they may incorporate all kinds of indirect effects from 
forcing a stationary state of a dynamics to be represented by a Bayesian belief network, potentially masking a far simpler underlying reality.

\section{Summary and discussion}
\label{sec:discussion}
We have given a brief introduction to causal analysis
and discussed how it extends the tools of factor graphs
and probabilistic models to describe outside interventions
that change the models themselves. We have compared and contrasted
causal analysis to the analysis of dynamic processes by
the dynamic cavity method and pointed out similarities
between causal analysis and physical correlation-response theory.

The \emph{do calculus} consideres causal relationships to be the fundamental building blocks of reality~\cite{Pearl-Causality}
and aims to discover which are these casual dependences in the system under investigation (usually DAG networks). 
Although this certainly is a fascinating goal, it is at odds with physical theory which 
does not admit causes and effects in the philosophical sense on the fundamental level, but
only for macroscopic (irreversible) processes. For such processes the flow of time is however essential,
and causes are thus naturally understood as variables influencing transition rates between various states in a system.
The stationary states of such processes are normally quite complicated reflecting not only
the the dependencies in the transition rates, but also chains of such dependencies of arbitrary length,
the only major exception being systems in thermodynamic equilibrium. 
Therefore, caution is needed when interpreting the results of causal analysis as causes in 
an everyday -- \textit{e.g} legal -- sense. The mechanisms identified by causal analysis include (except in thermal equilibrium)
both underlying direct effects and many kinds of indirect effects where the setting of one variable
influences the behaviour of another at a later time through one or many intermediaries.

The major advantage of causal analysis is instead in its relative simplicity of its basic ansatz.
Up to recent times few methods except Monte Carlo simulations were available to analyse the dynamics 
of non-equilibrium systems, and determining their stationary states was therefore laborious.  
We have discussed that for some systems the dynamic cavity offers an alternative approach
which could therefore be used as an alternative starting point of causal analysis.
Many major issues however remain to be solved in that approach, the most important
one perhaps being how to extend the dynamic cavity (if possible) to continuous-time processes.

\ack
E.A. acknowledges valuable discussions with 
Dr.~Antti Hyttinen and Profs.~Timo Koski and Jukka Corander.
This work has been supported by the European Union through 
Marie Curie ITN ``NETADIS'' (GDF), by Swedish Science Council through 
grant 621-2012-2982 (EA), and by the Academy of Finland through its Center 
of Excellence COIN (EA).


\end{document}